# A large-scale Twitter dataset for drug safety applications mined from publicly existing resources

Ramya Tekumalla
Department of Computer Science
Georgia State University
Atlanta GA USA
rtekumalla1@gsu.edu

Juan M Banda
Department of Computer Science
Georgia State University
Atlanta GA USA
jbanda@gsu.edu

**ABSTRACT**

With the increase in popularity of deep learning models for natural language processing (NLP) tasks, in the field of Pharmacovigilance, more specifically for the identification of Adverse Drug Reactions (ADRs), there is an inherent need for large-scale social-media datasets aimed at such tasks. With most researchers allocating large amounts of time to crawl Twitter or buying expensive pre-curated datasets, then manually annotating by humans, these approaches do not scale well as more and more data keeps flowing in Twitter. In this work we re-purpose a publicly available archived dataset of more than 9.4 billion Tweets with the objective of creating a very large dataset of drug usage-related tweets. Using existing manually curated datasets from the literature, we then validate our filtered tweets for relevance using machine learning methods, with the end result of a publicly available dataset of **1,181,993** million tweets for public use. We provide all code and detailed procedure on how to extract this dataset and the selected tweet ids for researchers to use.

## 1  Introduction

The World Health Organization (WHO) defined Pharmacovigilance as "the science and activities relating to the detection, assessment, understanding and prevention of adverse effects or any other drug-related problem" [1]. The aim of pharmacovigilance is to enhance patient care and safety in relation to the use of medicines; and to support public health programmes by providing reliable, balanced information for the effective assessment of the risk-benefit profile of medicines. Traditionally, clinical trials are employed to identify and assess the profile of medicines. However, since they have limited ability to detect all ADRs due to factors such as small sample sizes, relatively short duration, and the lack of diversity among study participants, post marketing surveillance is required [2]. The Food and Drug Administration (FDA) provides several post marketing surveillance programs like FDA Adverse Event Reporting System (FAERS), MedWatch to report events, however, under-reporting limits its effectiveness. A review of 37 studies established that more than 90% of ADRs are estimated to be under-reported [3]. Social media platforms like Twitter and Facebook contain an abundance of text data that can be utilized for pharmacovigilance [4]. Many studies presented satisfactory results by utilizing social media for pharmacovigilance and helped create a curated dataset for drug safety [5]. Several other studies presented a connection between drugs and addictive behavior among students using Twitter [6,7]. However, it is challenging to use Twitter due to limitations with service providers who can export only 50,000 tweets per day. Further, usage of Twitter's API or software to extract tweets is extremely time-consuming and economically unviable, especially for obtaining tweets relevant to a particular domain. Additionally, machine learning and deep learning models need exorbitant amounts of training data to train a model and not much data is available publicly for training.

In this context, data sharing [8] is a novel idea for research parasites to scavenge available datasets and apply their methodologies. Recent studies [9–11] prove that data sharing improves quality and strengthens research. The increase in collaborative efforts will provide an opportunity for researchers to continually enhance research ideas and avoid redundant research efforts [12,13]. As part of this research, we scavenged a large publicly available dataset and procured data related to pharmacovigilance. In this paper, we present a data corpus of 1,181,993 carefully filtered tweets. The deliverables [14,15] include filtered 1,181,993 tweet ids, code to download and separate tweets from the Internet Archive (IA). Due to Twitter's terms of service, tweet text cannot be shared. Therefore, only tweet ids are publicly made available. The whole methodology can be reproduced using the deliverables. This corpus can help train machine learning and deep learning models and can be reused by other researchers to enhance research in Pharmacovigilance.

## 2  Data Preparation

The Internet Archive (IA) [16] is a non-profit organization that builds digital libraries of Internet sites and other cultural artifacts in digital form and provides free access to researchers, historians and scholars. For this research, we used the largest publicly available Twitter dataset in Internet Archive, which contains several json files of tweets in tar files sorted by date for each month of the year. The tar file must be downloaded and decompressed before usage. A total of 9,406,233,418 (9.4 billion) tweets for the years 2012 to 2018 are available in this dataset. We filtered this data using a drug terms dictionary to identify drug-specific tweets. The time taken to download, process and filter these tweets was 132 days.

## 2   Drug Dictionary Creation

The UMLS [17] is a large, multi-purpose and multilingual vocabulary database that contains information about biomedical and health related concepts, their various names, and the relationships among them. The UMLS includes the Metathesaurus, the Semantic Network, and the SPECIALIST Lexicon and Lexical Tools. Metathesaurus, the biggest component of UMLS was utilized in creating the drug dictionary, more specifically the RxNorm [18] vocabulary. This vocabulary provides normalized names for clinical drugs and link names to the drug vocabulary commonly used in pharmacy management and drug interaction software. The MRCONSO table was filtered using RxNorm and Language of Term (LAT), which was set to English. The filtered table contained a total of 279,288 rows. Since, the dictionary was used on Twitter data and the total number of characters allowed in a tweet was 140 (until 2017) and 280 (from October 2017 onwards), we eliminated all the strings of length less than or equal to 3 (too ambiguous) and greater than or equal to 100.  This was due to a less likely chance for tweets to contain drug names that were as short as 3 characters or as long as 100 characters. Further, we removed strings such as "2,10,15,19,23-pentahydrosqualene" which are chemical compounds. This elimination was based on the premise that users would find it cumbersome and tedious to type detailed chemical names of drugs, especially on social media. Additionally, we removed 50 terms like "disk, foam, bar-soap, sledgehammer, cement, copper, sonata" as these terms are not commonly used as drug names and in pharmacovigilance. After deleting the common terms and chemical compounds, only 266,556 rows were available of which five term types were used in the drug dictionary for the research. The dictionary also consists of a Concept Unique Identifier (CUI) to which strings with the same meaning are linked. The CUI is used in order to ensure that the meanings are preserved over time regardless of the different terms that are used to express those meanings. All the strings have been converted to lowercase and trimmed of white spaces. A total of 111,518 unique strings were used in total to create the drug dictionary. Table 1 represents the number of strings used for each term type and Table 2 contains sample rows from the drug dictionary.

| Term Type | Example | # Strings |
|---|---|---|
| Ingredients (IN) | Fluoxetine | 11,427 |
| Semantic Clinical Drug Component (SCDC) | Fluoxetine 4 MG/ML | 27,038 |
| Semantic Branded Drug Component (SBDC) | Fluoxetine 4 MG/ML [Prozac] | 17,938 |
| Semantic Clinical Drug (SCD) | Fluoxetine 4 MG/ML Oral Solution | 35,112 |
| Semantic Branded Drug (SBD) | Fluoxetine 4 MG/ML Oral Solution [Prozac] | 20,003 |

Table 1: **Term types, their definitions and Number of strings**

| Concept Unique Identifier (CUI) | Term String |
|---|---|
| C0290795 | adderall |
| C0700899 | benadryl |
| C0025219 | melatonin |
| C0162373 | prozac |
| C0699142 | tylenol |

Table 2: **Sample drug dictionary**

## 3   Methods

In order to identify drug-specific tweets that would be useful for pharmacovigilance, we applied the drug dictionary on the Internet Archive twitter dataset. We filtered the dataset using spaCy, an open-source library in Python. We used the matcher in spaCy which would match sequences of tokens, based on pattern rules. Subsequently, the program generates an output file with the filtered tweets if it finds a match with the drug dictionary in the tweet text. Tweets are retrieved only if their language is set to English and if they are not retweeted. Retweets were not included because we did not want to add bias to the resulting dataset. If there are a number of tweets with the same text, then when we use the dataset to train the model, the model would be biased. Additionally, in pharmacovigilance any drug signals would be greatly amplified in potential error. Initially, the method was performed on 2018 data, with our results showing that the maximum number of tweets that got separated consisted of a single drug string (one term). We speculate that, since Twitter has a limitation on the number of characters, people tend to write abbreviated terms or single terms that are either drug names or ingredients, instead of a drug string that consists of 4 or 5 terms. We experimented using the dictionary which consists of 111,518 unique drug strings (Table 1) for all the tweets from 2018. After analyzing the retrieved tweets, the majority of the tweets contain only single terms (Example: "belatacept") and not multiple terms (Example: "belatacept 250 mg injection"). Therefore, we identified the single terms and created a single string dictionary. A total of 6 programs are run on each month for the year 2018. Only 10 months of data was available for the year 2018. Number of tweets obtained for four months for the year 2018 when used on six dictionaries are presented in table 3.

|  | Total | Single string | Ing. | SCD | SCDC |
|---|---|---|---|---|---|
| Jan. | 134,747,413 | 83,583 | 27,718 | 0 | 15 |
| Aug. | 141,132,076 | 67,227 | 21,335 | 0 | 13 |
| Sept. | 133,068,824 | 64,123 | 22,230 | 0 | 22 |
| Oct. | 132,297,280 | 68,221 | 22,955 | 0 | 17 |
| Total | 1,102,507,263 | 385,503 | 196,788 | 1 | 112 |

Table 3: **Number of tweets obtained for each month from the Internet Archive Dataset in 2018**

The SCD, SBD and SBDC dictionaries did not yield any tweets from 2018. In order to determine the reason, we examined and analyzed the dictionaries. For each term type in the drug dictionary, we calculated the lengths of all drug strings and identified the number of characters at each length ranging between 4 to 99 characters. Further, we also noted the average and median lengths of the drug strings. Table 4 depicts detailed statistics for each term type.

| Term Type | Minimum length (#) | Maximum length (#) | Average length |
|---|---|---|---|
| Ingredients | 4 (11) | 99 (1) | 21.556 |
| SCDC | 10 (16) | 93 (2) | 24.152 |
| SBDC | 19 (1) | 99 (54) | 43.757 |
| SCD | 20 (1) | 99 (99) | 48.531 |
| SBD | 32 (3) | 99 (84) | 57.818 |

Table 4: **Statistics for each drug term type**

SBD, SCD and SBDC had the highest number of lengthy drug strings. The following drug string from SCDC drug dictionary, "pneumococcal capsular polysaccharide type 33f vaccine 0.05 mg ml", has 64 characters. It is impractical to type the whole drug string in a tweet without an error. 90% of the tweets obtained from the SCDC were advertisements on either promoting the product or selling the product. Further examining all the tweets, we eliminated the 4 dictionaries (SCD, SCDC, SBD, SBDC) and used the single string and ingredients dictionary since it saves an enormous amount of computation time.

The following table represents the number of tweets retrieved for the years from 2012 to 2018 when used with the two remaining drug dictionaries.

| Month | Total tweets | Single string | Ingredients |
|---|---|---|---|
| 2018 | 1,102,507,263 | 588,854 | 26,034 |
| 2017 | 1,448,114,354 | 1,079,616 | 52,058 |
| 2016 | 1,427,468,805 | 1,252,057 | 62,514 |
| 2015 | 1,224,040,556 | 1,149,736 | 55,367 |
| 2014 | 1,086,859,898 | 873,937 | 47,953 |
| 2013 | 1,871,457,526 | 1,463,184 | 85,618 |
| 2012 | 1,245,785,016 | 76,795 | 4,139 |
| Total tweets | 9,406,233,418 | 6,152,862 | 314,680 |

Table 5: Number of single and ingredient tweets obtained from total tweets

We made the code publicly available for reproducibility. A total of 132 days were required to download, unzip and filter the tweets using the drug dictionary. For all the years, each month was downloaded individually, unzipped to retrieve the json file and then the tweets were filtered using the drug dictionary. Typically, for a month, the method would require 10 minutes to download, 5 hours to unzip and 2 days to filter tweets on an IBM Blade Server with 768 GB RAM, 2 x Intel® Xeon® E5-269880 Processors, with 40 cores each, and 12TB of hard disk space.

The single string and ingredients dictionary was used on the IA dataset and a total of 6,703,331 (6.7 million) tweets were retrieved from 9,406,233,418 (9.4 billion) tweets. After eliminating duplicate tweets, a total of 6,703,166 were retrieved. We examined the retrieved tweets and found that more than 50% of the tweets are not relevant to pharmacovigilance. This is because some drug strings are used in common terminology and in other fields like math, technology etc. For example, the drug string "tablet" was used in reference to the electronic gadgets (Samsung, Microsoft tablets). In order to eliminate the tweets that are relevant to other domains and not pharmacovigilance, we employed machine learning classification models to filter tweets.

## 4 Classification

Since the filtered tweets contain a number of irrelevant tweets, we experimented with several classical machine learning models on the filtered tweets to clean the tweets. The tweets obtained after classification can be used for training different machine learning and deep learning models by other researchers. Since there are no trainable datasets that we could make use of, we created a dataset utilizing annotated datasets from different sources [19–22]. We emphasize that we did not annotate or create any annotated set of tweets ourselves.

## 4.1 Classical Models

We collected 259,042 tweets that only have drug strings from multiple papers on pharmacovigilance using social media [19–22] and downloaded all the tweets available through them. These tweets were annotated by different annotators as part of their research. The collected 259,042 tweets from multiple pharmacovigilance papers were labelled as "drug" tweets. Additionally, we randomly collected 300,208 non-drug tweets from multiple years from Internet Archive and labelled them as "non-drug" tweets. Pre-processing was performed on the downloaded tweets by removing links and emojis and only tweet text was separated. A total of 559,250 tweets were used as an annotated training set, where only drug tweets were the actual annotated tweets collected from different sources. We experimented with five classifiers: Naive Bayes, Logistic Regression, Support Vector Machines (SVM), Random Forest and Decision Trees using the scikit-learn [23]. CountVectorizer() method, which essentially utilizes a bag of n-grams representation was used for all the Classical machine Models. Support-Vector Machine constructs a hyperplane or set of hyperplanes in a high- or infinite-dimensional space, which can be used for classification, regression, or other tasks like outliers detection. We used a LinearSVC which is similar to SVC, but implemented in terms of liblinear rather than libsvm, so it has more flexibility in the choice of penalties and loss functions and should scale better to large numbers of samples. Naive Bayes methods are a set of supervised learning algorithms based on applying Bayes' theorem with the "naive" assumption of conditional independence between every pair of features given the value of the class variable. We used the Multinomial Naive Bayes which implements the naive Bayes algorithm for multinomial distributed data and is one of the two classic naive Bayes variants used in text classification. A Random Forest is a meta estimator that fits a number of decision tree classifiers on various sub-samples of the dataset and uses averaging to improve the predictive accuracy and control over-fitting. The DecisionTreeClassifier uses a CART algorithm (Classification And Regression Tree). CART is a non-parametric decision tree learning technique that produces either classification or regression trees, depending on whether the dependent variable is categorical or numeric, respectively. However, the scikit uses an optimized version of the CART which does not support categorical values.

Each classifier model is applied on the stratified 75-25% (training - test) split of the annotated training set. We calculated precision, recall and F-score to evaluate each classifier and the results are tabulated in Table 6.

| Classifier | Precision | Recall | F-measure | Accuracy |
|---|---|---|---|---|
| Logistic Regression | 0.7535 | 0.7814 | 0.7672 | 0.8267 |
| Naive Bayes | 0.7106 | 0.8281 | 0.7649 | 0.8140 |
| *SVM* | *0.7773* | *0.8091* | *0.7929* | *0.8456* |
| Decision Tree | 0.7274 | 0.5120 | 0.6010 | 0.7516 |
| Random Forest | 0.7406 | 0.6814 | 0.7097 | 0.7963 |

Table 6: **Classification metrics for the classical machine learning models**

## 4.2 Calculating cut-off thresholds

We applied all classical machine learning models on the filtered 6 million tweets to predict the probability score of each tweet from Internet Archive dataset. In order to determine the most optimal probability cutoff, we applied mixture models concepts [24]. The way this methodology works is by taking all the probability scores and divide them into several hundreds of bins. A histogram of probability frequency is determined by calculating the number of observations in each bin. Based on hypothesis of mixture models, probability scores are distributed according to a mixture of two Gaussian distributions (drug and non-drug tweets). Finally, two highest peaks of two Gaussian distributions and one valley with most depth between the two peaks are detected and the valley's deepest point is used as cut-off point (threshold). In Figure 1, we plot the number of tweets that have a given probability score. Starting from an assigned probability of one, we cumulatively count the number of tweets we would keep at any given probability threshold. The plots visualizes the selectivity of each model and the number of tweets at each threshold limit. Note that the optimal cutoff threshold is displayed next to the model name in the figure legend.

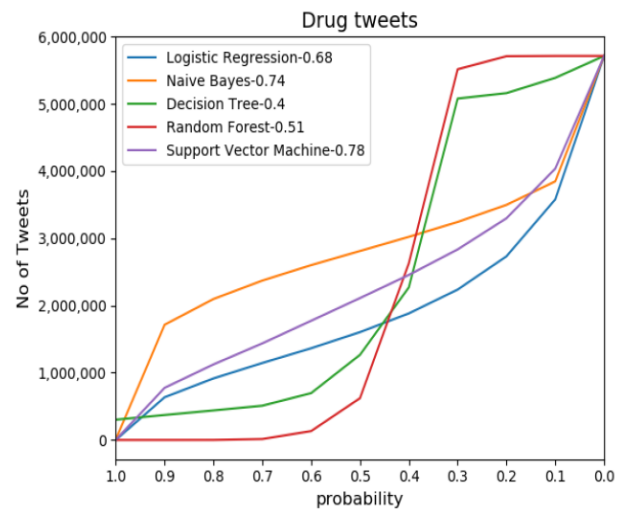

Figure 1: **Probabilities of drug tweets using Classical Models.**

Based on Tables 6 and the cutoff Figures 1, we selected SVM as the model to use to classify the relevance of the tweets. This model performed the best in terms of F-measure and accuracy, two of the metrics we deemed most relevant to identify useful tweets. After the classification filtering, we examined all the retrieved tweets and calculated the drug occurrences. We identified 6,867 unique drug strings in 1,181,993 million tweets. Figure 2 depicts the popular drug strings and the number of occurrences for each drug string in the classified tweets.

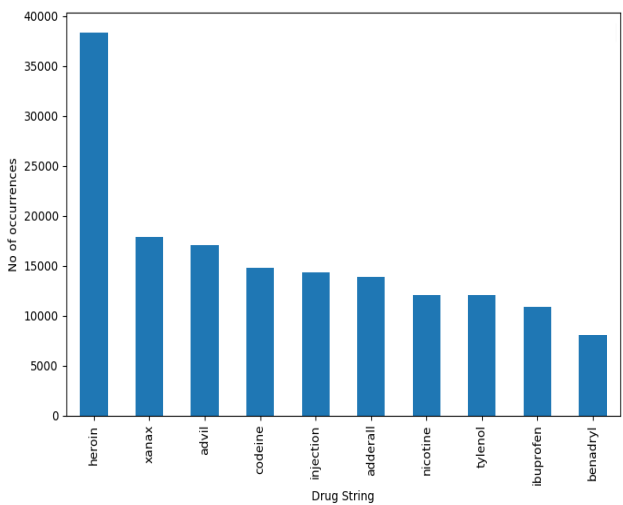

Figure 2: Popular drug string occurrences after filtering and classification of tweets.

Cocaine was the most popular drug string with 73,596 occurrences, however we eliminated it from the plot since it was used more as a recreational drug than a medical drug. The following are a few examples of the tweets. The drug strings are highlighted in bold.

1. "i need greasy food water and **tylenol** right now"

2. "i need some **morphine** for this pain"

3. "new year , new cough . drinking tropicana **vitamin c** & **zinc**. this stuff helps support a healthy immune system.".

4. "side effects of **memantine** #alzheimers #dementia".

5. "i took **advil** for the headache lol i don t have anything for sinuses."

## 5 Validation

In order to determine the quality of the dataset, we further conducted experiments to validate the created dataset. We designed two, bi modal classification experiments using Classical Machine Learning models since the computation time is faster when compared to the Deep Learning models. The training data includes tweets from the created dataset which were labelled as "drug tweets". Additional tweets were collected from twitter which did not contain drug terms and the tweets were labelled as "non-drug tweets". The test set consists of annotated tweets from publicly available resources [20,21] which were labelled as "drug tweets" and "non-drug tweets" collected from Twitter. Table 7 summarizes the different experiments the train and test data for the three experiments.

| Experiment Number | Train data | Test data |
|---|---|---|
| 1 | 2,000,000 tweets | 14,828 tweets |
| 2 | 134,063,321 tweets | 14,828 tweets |

Table 7: Details of the Experiments used for validation

Experiment 1 utilizes a total of 2 million tweets, out of which the top 1 million tweets were obtained from 1,181,993 tweets and were labelled as drug tweets. Additionally 1 million tweets obtained from the Internet Archive were labelled as "non drug tweets". Experiment 2 consists of 12,978,348 tweets from which 6,703,166 tweets were considered as "drug tweets". These are all the tweets obtained from the Internet Archive using the drug dictionary before applying any machine learning models. An additional 6,703,166 "non-drug" tweets were obtained from the Internet Archive. All the experiments use the same test data which consists of 7,414 annotated tweets labelled as "drug tweets" obtained from publicly available resources and 7,414 tweets collected from Twitter labelled as "non-drug" tweet. A tweet would classify as a "non-drug" tweet if and only if a tweet does not include a drug term. For each experiment, 5 classical machine learning models were employed. The classical machine learning models include Naive Bayes, SVM, Random Forest, Decision Tree and Logistic Regression. Table 8 summarizes the best results for each experiment.

| # | Best Model | Precision | Recall | F-measure | Accuracy |
|---|---|---|---|---|---|
| 1 | SVM | 0.9964 | 0.8331 | 0.9075 | 0.9150 |
| 2 | *SVM* | *0.9982* | *0.8369* | *0.9104* | *0.9177* |

Table 8: Best Model results obtained from validation

The SVM model obtained the best results when compared to the other models for all the experiments. The results from the experiments demonstrate that the dataset can be used to train machine learning models for research in Pharmacovigilance. It would be valuable to manually annotate a small set for additional validation, however, we believe that we can use ML and noisy labels (keyword searches), at scale, to increase the number of available datasets using 'weak supervision'. This in turn will avoid the need for additional manual curation of datasets and use weak supervision to produce silver standard datasets that are larger, as we show in our validation of the models built using this noisy data on existing manually annotated datasets.

## 6  Future Work

The proposition of this paper is to utilize publicly available resources and employ Machine and Deep Learning techniques to create a dataset that can be made available for pharmacovigilance research. We believe that we can't keep training models with very limited amount of manually annotated tweets, but we can use the theory of noisy labeling to create more robust models with silver standards [25-27]. However, there are a few limitations which we would like to address in our future work. This research utilizes only English tweets since there were no publicly available annotated drug tweets for other languages. We have not employed a spelling correction module in any of our experiments. This eliminates all the tweets with incorrect drug spellings. In future, we would like to employ a spelling check module, where we include all the incorrect spellings for the drugs which would increase the scope of the dataset. Furthermore, the annotated drug tweets used in the training data were collected from publicly available sources and are labelled as drug tweets. Hence, edge cases such as ambiguous tweets were not considered. In addition, we would like to experiment with Deep Learning models and obtain a model that can identify drug related tweets. In the future, we would like to develop an improved annotated dataset which can be utilized as a gold standard dataset, following a tri-modal distribution of probabilities where the edge cases are considered.

## 7  Conclusion

In this paper, we scavenged a publicly available Twitter dataset, Internet Archive, mining over 9.4 billion tweets. Using a simple drug dictionary and plenty of computing power, we filtered 6 million tweets with relevant drug terms in them. In order to determine the viability of the filtered tweets for research work, we used publicly available, manually and expertly curated tweet datasets to build classification models to identify the relevant (or similar) tweets in our dataset. Further, to determine the quality of the dataset, we created different experiments and achieved results which proves the worth of the dataset. Overall, these tasks took around 150 days for downloading, filtering and classification, in order to retrieve 1,181,993 tweets which can be used for drug safety research and as a training set for other supervised methods by researchers. We believe that, rather than using expensive manually curated datasets, we can use a heuristic (curated keyword search) to generate enough noisy labels and then use machine learning to further refine them to perform well against existing curated datasets. This methodology allows us to have larger datasets that could be used with modern machine learning methods that need more data to build 'better' models. We do show an evaluation to show that models built with these 'noisy labels' perform well in finding actually manually curated sets of tweets for the same purpose. Further, we believe that this approach can be reused and extended to several other domains by changing the dictionaries and the filtering mechanisms